\documentclass[
 reprint,
 amsmath,amssymb,
 nofootinbib,
 aps,
 prd,
 superscriptaddress,
 floatfix
]{revtex4-2}
\usepackage{aas_macros}
\usepackage{graphicx}
\usepackage{dcolumn}
\usepackage{bm}
\usepackage{amsmath}
\usepackage{float}
\usepackage{lipsum}
\usepackage{xcolor}
\usepackage{times}
\usepackage{textcomp}
\usepackage{latexsym}
\usepackage[hidelinks]{hyperref}
\usepackage{acro}
\usepackage{mathtools}
\usepackage{url}
\usepackage{aas_macros}
\usepackage[utf8]{inputenc}
\usepackage[normalem]{ulem}
\usepackage{xspace}
\usepackage{lineno}
\usepackage{scrextend}
\definecolor{lightgreen}{RGB}{179,255,164}
\definecolor{lightblue}{RGB}{196,206,255}
\definecolor{light-gray}{RGB}{230,230,230}

\newcommand{\hbos}{\ensuremath{\mathcal{H}_{\mathrm{B}}}\xspace}
\newcommand{\hnull}{\ensuremath{\mathcal{H}_{\rm {A}}}\xspace}

\newcommand{\NruleoutFlat}{$30^{+135}_{-25}$}
\newcommand{\NconfirmHeavyFlat}{$65^{+165}_{-55}$}
\newcommand{\NconfirmLightFlat}{$25^{+95}_{-15}$}
\newcommand{\NruleoutLow}{$140^{+120}_{-105}$}
\newcommand{\NconfirmHeavyLow}{$155^{+345}_{-145}$}
\newcommand{\NconfirmLightLow}{$80^{+210}_{-70}$}
\begin{document}
\title{Searching for ultralight bosons with spin measurements of a population of binary black hole mergers}
\author{Ken~K.~Y.~Ng}
\email{kenkyng@mit.edu}
 \affiliation{LIGO Lab, Department of Physics, and Kavli Institute for Astrophysics and Space Research, Massachusetts Institute of Technology, 77 Massachusetts Avenue, Cambridge, Massachusetts 02139, USA}
\author{Otto~A.~Hannuksela}
\email{o.hannuksela@nikhef.nl}
\affiliation{Nikhef -- National Institute for Subatomic Physics, Science Park, 1098 XG Amsterdam, The Netherlands}
\affiliation{Department of Physics, Utrecht University, Princetonplein 1, 3584 CC Utrecht, The Netherlands}
\affiliation{Department of Physics, The Chinese University of Hong Kong, Shatin, NT, Hong Kong}
\author{Salvatore Vitale}
 \affiliation{LIGO Lab, Department of Physics, and Kavli Institute for Astrophysics and Space Research, Massachusetts Institute of Technology, 77 Massachusetts Avenue, Cambridge, Massachusetts 02139, USA}
\author{Tjonnie~G.~F.~Li}
 \affiliation{Department of Physics, The Chinese University of Hong Kong, Shatin, NT, Hong Kong}
 \affiliation{Institute for Theoretical Physics, KU Leuven, Celestijnenlaan 200D, B-3001 Leuven, Belgium}
 \affiliation{Department of Electrical Engineering (ESAT), KU Leuven, Kasteelpark Arenberg 10, B-3001 Leuven, Belgium}
\date{\today}

\begin{abstract}
Ultralight bosons can form clouds around rotating black holes if their Compton wavelength is comparable to the black hole size. The boson cloud spins down the black hole through a process called superradiance, lowering the black hole spin to a characteristic spin determined by the boson mass and the black hole mass.
It has been suggested that spin measurements of the black holes detected by ground-based gravitational-wave detectors can be used to constrain the mass of ultralight bosons.
Unfortunately, a measurement of the \emph{individual} black hole spins is often uncertain, resulting in inconclusive results.
Instead, we use hierarchical Bayesian inference to \emph{combine} information from multiple gravitational-wave sources and to obtain stronger constraints.
We show that hundreds of high signal-to-noise ratio gravitational-wave detections are enough to exclude (confirm) the existence of noninteracting bosons in the mass range $\left[10^{-13},3\times 10^{-12}\right]$~eV $\left([10^{-13},10^{-12}]~\rm{eV}\right)$.
The precise number depends on the distribution of black hole spins at formation and the mass of the boson.
\end{abstract}

\maketitle

\section{Introduction}\label{sec:intro}
Ultralight bosons with masses $\lesssim10^{-11}\,\mathrm{eV}$, including axionlike particles~\cite{Peccei:1977hh,Weinberg:1977ma,Wilczek:1977pj}, dilatons and, moduli~\cite{Dimopoulos:1996kp,Goodsell:2009xc,Arvanitaki:2009fg} and fuzzy dark matter~\cite{Preskill:1982cy,Abbott:1982af,Dine:1982ah,Turner:1990uz,Hu:2000ke}, have been proposed as a potential solution to various problems ranging from fundamental physics to cosmology~\cite{Peccei:1977ur,Weinberg:1977ma,Wilczek:1977pj,Peccei:2006as,Bertone:2004pz,Hertzberg:2008wr,Arvanitaki:2009fg,Jaeckel:2010ni,Arvanitaki:2010sy,Essig:2013lka,Marsh:2015xka,Hui:2016ltb,Arvanitaki:2019rax}.
Efforts are underway to search for these ultralight bosons using table-top experiments or astronomical observations~\cite{Asztalos:2009yp,Wagner:2010mi,Rybka:2010ah,Arik:2011rx,Pugnat:2013dha,Arvanitaki:2014faa,Corasaniti:2016epp,Choi:2017hjy,Akerib:2017uem,Brubaker:2016ktl,Kim:2017yen,Garcon:2017ixh,Arvanitaki:2014wva,Arvanitaki:2016qwi,Baryakhtar:2017ngi,Brito:2017wnc,Cardoso:2018tly,antonio2018cw,Ghosh:2018gaw,Tsukada:2018mbp,Stott:2017hvl,Hannuksela:2018izj,Ouellet:2018beu,Davoudiasl:2019nlo,Fernandez:2019qbj,Palomba:2019vxe,Ng:2020jqd,Abel:2017rtm,Grote:2019uvn,Dev:2016hxv,Zhu:2020tht,Ng:2020ruv}.
When the Compton wavelength of the hypothetical boson is comparable to the size of a black hole (BH), i.e., $\alpha \equiv GM\mu_s/\hbar c^3 \sim1$, where $\alpha$ is the ``gravitational fine-structure constant'', $M$ is the BH mass and $\mu_s$ is the boson mass, then a classical wave amplification process (\emph{superradiance}) forms a bosonic cloud around the BH~\cite{1971JETPL..14..180Z,1972JETP...35.1085Z,Press:1972zz,Misner:1972kx,Dolan:2007mj,Arvanitaki:2009fg,Arvanitaki:2010sy,Brito:2014wla,Brito:2015oca}.
The formation of this cloud extracts rotational energy from the BH until the BH reaches a critical spin set by the Compton frequency of the boson and the mass of the BH~\cite{Dolan:2007mj,Arvanitaki:2009fg,Arvanitaki:2010sy,Brito:2014wla,Brito:2015oca}.
This results on a critical spin curve on the BH mass-spin plane (e.g. Fig.~3 of Ref~\cite{Arvanitaki:2010sy}):
BHs born with spins above the critical spin curve will rapidly spin down until their final masses and spins lie on the curve. The net result of the superradiance process is thus to carve a region of the mass-spin plane, above the critical spin curve, where BHs are unlikely to be observed (``exclusion region'').
Since the exact position and extent of the exclusion region depend on the boson mass (again, Fig.~3 of Ref~\cite{Arvanitaki:2010sy}), one can use mass and spin measurements for a population of BHs to search and characterize ultralight bosons~(e.g.~\cite{Arvanitaki:2009fg,Arvanitaki:2010sy,Arvanitaki:2014wva,Arvanitaki:2016qwi,Baryakhtar:2017ngi,Fernandez:2019qbj}).

Spin measurements of BHs in X-ray binaries (see e.g. References~\cite{Remillard:2006fc,Middleton:2015osa}) could be used to search for bosons with their masses commensurate with BHs in the mass range $5~\mathrm{M}_{\odot}< M\lesssim 20~\mathrm{M}_{\odot}$~\cite{Remillard:2006fc,Corral-Santana:2015fud}. Gravitational-wave (GW) signals emitted by binary black holes (BBHs) provide another avenue, as they encode the properties of their sources, including the masses and spins of the two component BHs. Since ground-based GW detectors such as LIGO and Virgo~\cite{aligo,virgo} can detect heavier BHs ($M$ up to $\sim100~\rm{M}_{\odot})$~\cite{GWTC1,GWTC2} than those found in X-ray binaries, the spin measurements inferred from GWs probe a lighter range of boson mass.
This probe is complicated by two facts: (i) the measurements of individual BH spins with ground-based GW detectors are usually challenging~\cite{vitale2014spinmagpe,Purrer:2015nkh,vitale2017spinmagpe,Fairhurst:2019vut} and
(ii) what is measured is the distribution of spins at merger, which is not only impacted by an eventual interaction with the bosons, but also on the distribution of spins at birth~\cite{Belczynski:2017gds,gerosa2018spins,postnov2019spins,Bavera:2019fkg}.

In this paper we perform hierarchical Bayesian inference on a {population} of BBHs, to \emph{simultaneously} infer the boson mass and the BH spin distribution at \textit{formation}~\cite{farr2017distinguishing,Roulet:2018jbe,Thrane:2018qnx,taylor2018hba,LIGOScientific:2018jsj,gaebel2019gwpop,mandel2019gwhba,Vitale:2020aaz,Roulet:2020wyq}.
First, we show that the existence of ultralight bosons in the mass range between $10^{-13}$~eV and $3\times10^{-12}~\mathrm{eV}$ can be ruled out with $\mathcal{O}(100)$ high signal-to-noise ratio (SNR) BBH detections.~\footnote{In the paper, we define ``high SNR'' as SNR$\geq30$.}
Second, we illustrate how our method can confirm the existence of bosons with two examples of boson masses, $\mu_s=10^{-12}$ and $10^{-13}$~eV.

\section{Critical spin arising from superradiant instability}\label{sec:criticalspin}
GW measurements yield the masses and spins of BHs at the time of merger.
Therefore, one needs to account for the impact of superradiance on the evolution of spins, which we quickly review here.
Superradiance causes the growth of a boson cloud in a time scale $\tau^{\rm inst}_{[nlm]}$, called \emph{instability time scale}, and eventually spins down the host BH to a characteristic critical spin $\chi^{\infty}_{[nlm]}$ (which will be defined below)~\cite{Dolan:2007mj,Arvanitaki:2009fg,Arvanitaki:2010sy,Brito:2015oca}.
The indices $[nlm]$ are the analog of the hydrogen atom quantum numbers, i.e., a set of radial, orbital azimuthal, and magnetic quantum numbers, of the cloud's bound states.~\footnote{We follow Dolan's notation, in which the ordinary principal quantum number $\tilde{n}=n+l+m$, such that $n=0$ corresponds to the dominant (nodeless) modes~\cite{Dolan:2007mj}.}
For any given $[nlm]$ mode, the instability time scale is representative of the exponential growth of the occupation number of the bosons in that mode, such that modes with small $\tau^{\rm inst}_{[nlm]}$ are populated quickly.

One often introduces the inverse of the instability time scale $\tau^{\mathrm{inst}}_{[nlm]}$, called the \emph{superradiant rate} $\Gamma^{\mathrm{inst}}_{[nlm]}$, which can be analytically calculated for $\alpha \ll 1$~\cite{Detweiler:1980uk,Arvanitaki:2009fg,Arvanitaki:2010sy,Brito:2014wla,Brito:2015oca} (hereafter, we set $G=c=\hbar=1$),

\begin{align}\label{eq:tauinst}
 \begin{split}
  \Gamma^{\mathrm{inst}}_{[nlm]} = & \mu_s (\mu_s M)^{4l+4} (m \chi - 2 \mu_s r_+) \\
  & \times \frac{2^{4l+2}(2l+1+n)!}{(l+1+n)^{2l+4}n!}\left[ \frac{l!}{(2l)!(2l+1)!} \right]^2  \\
  & \times \prod_{k=1}^l [k^2 (1-\chi^2)+(m \chi -2 \mu_s r_+)^2],
 \end{split}
\end{align}
where $\chi$ is the dimensionless spin of the BH 
and $r_+\equiv M\left( 1+\sqrt{1-\chi^2} \right)$ is the radial coordinate of the BH outer horizon.
For the first few $l$'s, the fastest growing mode is the one with $l=m$, such that $\Gamma^{\mathrm{inst}}_{[nlm]}$ is the largest~\cite{Arvanitaki:2010sy}.
For any given $M$ and $\mu_s$, superradiance can happen as long as $\Gamma_{[nlm]}^{\mathrm{inst}}(\mu_s,M,\chi)>0$; hence,

\begin{align}\label{eq:SRconditions}
    \frac{\alpha}{m}<\frac{\chi}{2\left(1+\sqrt{1-\chi^2}\right)}<\frac{1}{2}.
\end{align}
which gives a condition on the BH spin $\chi$ such that superradiance can happen for the mode $[nlm]$.
If the BH-cloud system had an infinite amount of time to evolve without disturbances, the spin angular momentum of the BH would eventually be lowered to the point where Eq.~(\ref{eq:SRconditions}) cannot be satisfied or, equivalently, where $\Gamma_{[nlm]}^{\mathrm{inst}}=0$ (saturation of the superradiance).
The \textit{critical} spin for the $[nlm]$ mode is thus defined as the spin below which the superradiant growth of the cloud of the $[nlm]$ mode is forbidden,

\begin{align}\label{eq:criticalSpin}
    \chi^{\infty}_{[nlm]}=\frac{4\alpha m}{4\alpha^2 + m^2}.
\end{align}

Besides the spin angular momentum of the BH, a small fraction $(\lesssim 10\%)$ of BH mass is also extracted and contributes to the cloud's mass~\cite{Arvanitaki:2010sy,Brito:2014wla,Ficarra:2018rfu}.
This is smaller or at most comparable to the BH mass uncertainty from GW measurements~\cite{vitale2017spinmagpe}.
Therefore, we neglect the BH mass loss due to superradiance and assume that the BH masses at merger are the same as the masses at formation.

In this study, we do not allow for boson self-interaction, which would lead to additional phenomenology~\cite{Arvanitaki:2010sy,Arvanitaki:2014wva,Baryakhtar:2020gao}. We refer to Ref.~\cite{Fernandez:2019qbj} for the analysis on self-interacting bosons using spin measurements of x-ray binaries.

\section{Postsuperradiant spin of astrophysical black holes}\label{sec:postSRspin}

Astrophysical BHs in binaries have finite lifetimes, $\tau_s$, which implies their spins at merger will not reach the critical spin $\chi^{\infty}_{[nlm]}$. 
Therefore, we need to calculate the postsuperradiant BH spin for the $[nlm]$ mode, $\chi_{[nlm]}$~\footnote{It is larger than $\chi^{\infty}_{[nlm]}$.} by solving $\Gamma^{\rm inst}_{[nlm]}(\mu_s, M, \chi_{[nlm]})=1/\tau_s$ for $\chi_{[nlm]}$, i.e., we truncate the spin evolution when the instability time scale decreases to the BH lifetime.

Naturally, if the lifetime of a BH is too short compared to the time required for the boson cloud to spin down its host BH, the effect might not even be measurable. Refs.~\cite{Arvanitaki:2010sy,Arvanitaki:2014wva,Arvanitaki:2016qwi} find that the bosonic field in the cloud should increase by $\sim 180$ $e$-foldings for the cloud to store a large amount of the BH angular momentum.
This $e$-folding requirement translates to a ``growth time scale'' for the boson cloud such that it can significantly spin down the BH, $\tau^{\rm grow}_{[nlm]}\approx 180~\tau^{\rm inst}_{[nlm]}(\mu_s,M,\chi_{I[nlm]})$, where $\chi_{I[nlm]}$ is the dimensionless spin of the BH at the onset of superradiance of the $[nlm]$ mode.
Therefore, a BH that is born with a spin at formation $\chi_F$ and merges in $\tau_s$ can be spun down to the postsuperradiant spin $\chi_{[nlm]}$ (which is given by the solution of $\Gamma^{\rm inst}_{[nlm]}(\mu_s, M, \chi_{[nlm]})=1/\tau_s$) only if $\chi_F>\chi_{[nlm]}$ and $\tau_s>\tau^{\rm growth}_{[nlm]}$.

\begin{figure}[ht]
    \centering
    \includegraphics[width=0.48\textwidth]{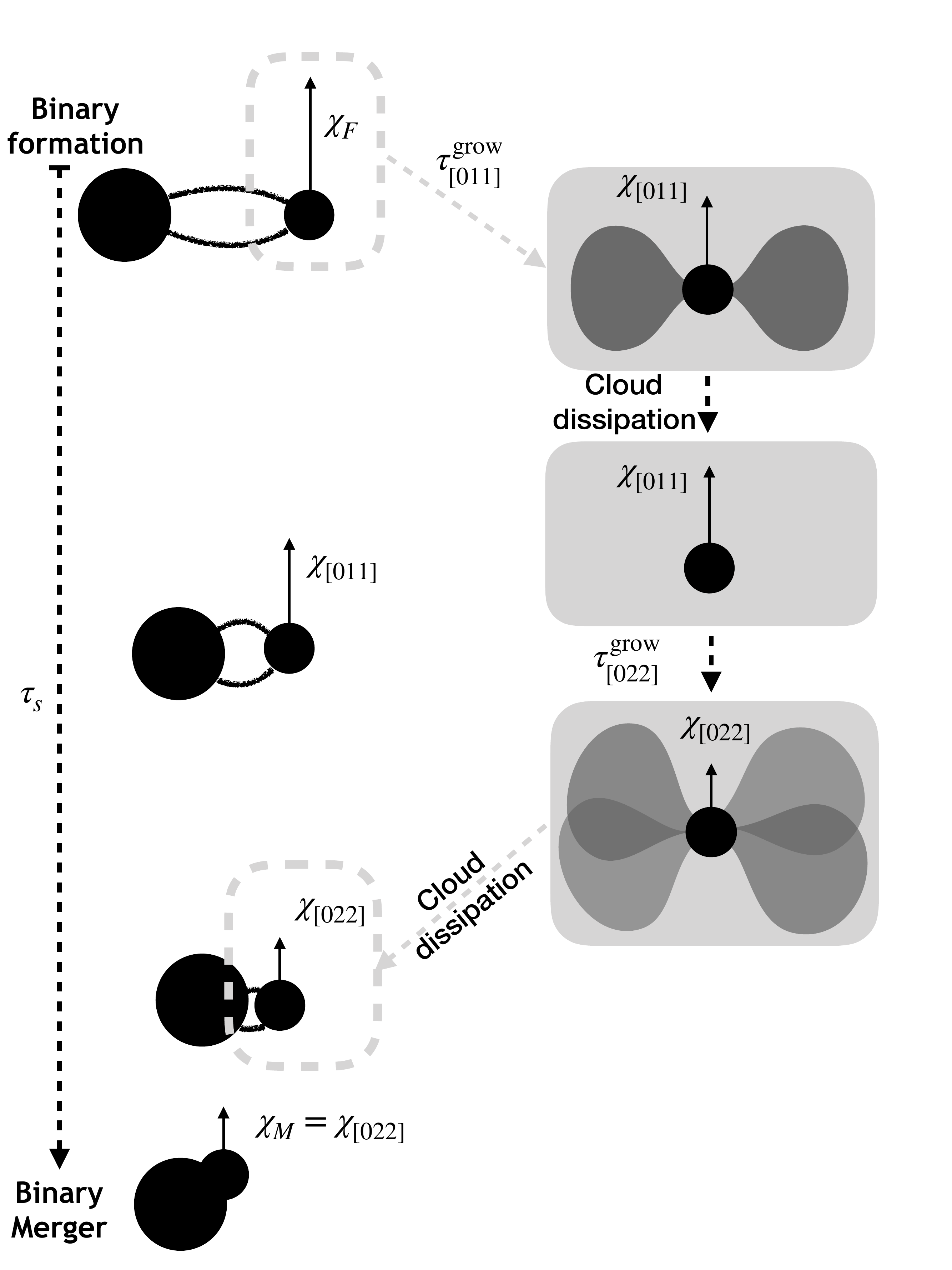}
    \caption{The possible evolution of a BBH system from its formation to merger. The cartoon follows the growth and dissipation of boson clouds around one of the BHs. The BH starts with a spin at formation $\chi_F$ large enough to trigger superradiant instability. The $n=0,l=m=1$ cloud forms with a characteristic time scale $\tau^{\rm grow}_{[011]}$, lowering the BH spin to $\chi_{[011]}$. The cloud is dissipated through monochromatic GW emission. In this example, the $\chi_{[011]}$ is large enough that a second cloud ($n=0,l=m=2$) is also created, and further reduces the BH spin to $\chi_{[022]}$. At this point the spin of the BH is too small to trigger the formation of higher-order clouds, and the BH will merge with spin $\chi_M=\chi_{[022]}$. The highest level of the cloud that can be populated depends on $\chi_F$ and on the inspiral time scale $\tau_s$.
    (The plot is not to scale.)}
    \label{Fig:Cartoon}
\end{figure}

Multiple clouds with different modes can be excited within the lifetime of a BH.
The highest mode of the cloud that can be populated is given by the condition $\tau^{\rm grow}_{[nlm]}~{<}~\tau_s~{<}~\tau^{\rm grow}_{[n(l+1)(m+1)]}$.
One can then estimate the BH spin at merger $\chi_M$ to be the postsuperradiant spin of this highest mode $\chi_{[nlm]}$, given the BH spin at formation and its lifetime.
For example, if a BH is born with $\chi_F>\chi_{[011]}$, then the formation of the initial cloud (with the time scale $\tau^{\rm grow}_{[011]}$) slows down the BH spin to $\chi_{[011]}$.
After its formation, the cloud dissipates away emitting nearly monochromatic gravitational waves~\cite{Arvanitaki:2010sy,Arvanitaki:2016qwi,Baryakhtar:2017ngi,Brito:2017wnc,Brito:2017zvb}.
Next, a second  (``higher mode'') cloud is formed with the time scale $\tau^{\rm grow}_{[022]}$, and the BH spins down further to the postsuperradiant spin of this next mode $\chi_{[022]}$, if $\chi_{[011]}>\chi_{[022]}$.
This cycle repeats until the BBH merges at time $\tau_s$.
In Fig.~\ref{Fig:Cartoon}, we show a schematic picture of a system for which both the $l=m=1$ and the $l=m=2$ clouds have enough time to form before merger.
We note that only the first few growing modes $n=0$ and $l=m\leq3$ are relevant for the typical astrophysical time scales $\tau_s\lesssim 10$~Gyr.

In principle, the BBH merger could cause the cloud to fall back to its host through level mixing~\cite{Arvanitaki:2010sy,Ficarra:2018rfu,baumann2019probing,berti2019ultralight,Baumann:2019ztm}.
While one could think that this results in a transfer of the cloud's angular momentum back to the host BH, which would then be spun-up, most of the in-falling modes have nonpositive angular momentum (i.e. $m\leq 0$) due to selection rules.~\cite{baumann2019probing,berti2019ultralight,Baumann:2019ztm}.
This is why recent studies have suggested that the in-falling cloud instead \emph{spins down} the host BH, or might decrease the orbital angular momentum in the binary system~\cite{Ficarra:2018rfu,baumann2019probing,berti2019ultralight,zhang2019levelmixing,zhang2019cloudbbh,Baumann:2019ztm}.
An eventual decrease in the BH spin by the fallback would further increase the size of the exclusion region on the BH mass-spin plane, making it easier to verify the existence of bosons with our method.
To be conservative, we ignore this binary effect on the postsuperradiant spins.

\section{Testing the Ultralight-boson Hypothesis using Hierarchical Bayesian inference}\label{sec:method}

An astrophysical distribution of BH spins at birth, which produces mainly small spins in the absence of superradiance, is partially degenerate with the postsuperradiant spin distribution that originates from a moderate (or high) spin distribution at birth in the presence of superradiance.~\footnote{Although in the latter case one would expect a characteristic peak at around the postsuperradiant spin curve.}
Hence, we need to simultaneously infer the spin distribution at birth and the boson mass to properly account for this degeneracy. 

We use hierarchical Bayesian inference~\cite{taylor2018hba,Thrane:2018qnx,gaebel2019gwpop,mandel2019gwhba,Vitale:2020aaz}, and consider two competing models: (i) in the ``boson model,'' \hbos, we assume that a boson exists such that BHs can spin down to the corresponding postsuperradiant spins $\chi_{[nlm]}$ through superradiance (Sec.~\ref{sec:postSRspin}); (ii) in the ``astrophysical model,'' \hnull, ultralight bosons do not exist, and the spins of BHs merging in binaries are entirely determined by their astrophysical evolution.

The two hypotheses are distinguishable through the resulting distribution of the BH spins at merger. Specifically, for \hbos, we assume
\begin{equation}
\hbos: \chi_M =\label{eq:HBSpins}
\begin{cases}
 \chi_{[nlm]}, &\text{if }\tau^{\rm grow}_{[nlm]}<\tau_s<\tau^{\rm grow}_{[n(l+1)(m+1)]}\\
 \chi_F, &\text{otherwise}
\end{cases}
\end{equation}
where $\chi_F$ and $\chi_M$ are the individual BH spins at formation and at merger, respectively, and $\tau_s$ is the BH lifetime.
The condition $\tau^{\rm grow}_{[nlm]}<\tau_s<\tau^{\rm grow}_{[n(l+1)(m+1)]}$ implies that $\chi_{[nlm]}$ is obtained as the postsuperradiance spin of the highest mode that can be populated within the binary lifetime.
If superradiance does not happen for the specific BH (either because of its parameters at formation or because of the time scale to merger) then the spin at merger is the same as the spin at formation, i.e. $\chi_M=\chi_F$.
On the other hand, there is no superradiance and therefore no spin evolution for \hnull.
Thus, for all black holes, the spin remains unchanged from formation to merger,
\begin{equation}
	\hnull: \chi_M=\chi_F.
\end{equation}

For both models, we parametrize the distribution of BH spins at formation with a beta distribution, controlled by two unknown shape parameters $\alpha>0$ and $\beta>0$: $p(\chi_F | \alpha,\beta)\propto \chi_F^{\alpha-1}(1-\chi_F)^{\beta-1}$. This is a generic functional form that can capture multiple different formation pathways~\cite{wysocki2018hba,LIGOScientific:2018jsj}.
The boson model thus depends on three hyperparameters $\Lambda_{\hbos}=(\alpha,\beta,\mu_s)$, while the astrophysical model only has two hyperparameters $\Lambda_{\hnull}=(\alpha,\beta)$.
We aim at measuring the hyperparameters $\Lambda$, given a set of $N$ GW observations $\mathbf{d}=\left\{ d_k \right\}$, whose morphology depends on a set of parameters $\theta$, such as BH masses, spins and distance~\cite{GW150914PE}.
The distribution of measured $\Lambda$, known as hyperposterior, can be written as~\cite{taylor2018hba,Thrane:2018qnx,gaebel2019gwpop,mandel2019gwhba,Vitale:2020aaz}:
\begin{equation}
p(\Lambda | \mathbf{d}) \propto \pi(\Lambda)\prod_{k}^{N} \left[ \frac{1}{\Sigma(\Lambda)} \int p(\theta | \Lambda) p(d_k | \theta) d \theta \right]
\end{equation}
where $p(\theta | \Lambda)$ is the expected distribution of the individual events parameters, given the hyperparameters; $ \pi(\Lambda)$ are the priors of the hyperparameters; $p(d_k | \theta)$ is the likelihood of the $k$th GW source; and $\Sigma(\Lambda)$ is the normalization factor given by

\begin{equation}
\Sigma(\Lambda) = \int p(\theta | \Lambda) p_{\mathrm{det}}(\theta) d\theta,\nonumber
\end{equation}
where $p_{\mathrm{det}}(\theta)$ is the detection probability for a BBH with parameters $\theta$.
The normalization factor $\Sigma(\Lambda)$ can thus be interpreted as the fraction of detectable BBHs.

When working with the boson model, $\Lambda_{\hbos}=(\alpha,\beta,\mu_s)$ and $\theta_{\hbos}=(M_1,M_2,\chi_{M,1},\chi_{M,2},\tau_s)$, where $M_i$ and $\chi_{M,i}$ are the masses and spins (at merger) of the two compact objects in the binary.
One thus has:

\begin{widetext}
\begin{equation}\label{eq:hyperpos}
	p\left(\Lambda_{\hbos} | \mathbf{d}, \hbos \right) \propto \pi(\Lambda_{\hbos})\prod_k^N \left\{ \frac{1}{\Sigma(\Lambda_{\hbos})} \int p(d_k | \theta_{\hbos}) \pi(M_1,M_2,\tau_s) \prod_{i=1}^2 \left[ p(\chi_{M,i} | \Lambda_{\hbos},M_i, \tau_s) dM_i d\chi_{M,i}\right] d\tau_s\right\} ,
\end{equation}
\end{widetext}

In this expression, $\pi(M_1,M_2,\tau_s)$ is the prior on the component masses and the merger time of BBHs, $\pi(\Lambda_{\hbos})$ is the prior on the hyperparameters of the model $\hbos$, and $p(\chi_{M,i}| \Lambda_{\hbos},M_i,\tau_s)$ is the distribution of the spin magnitude at merger of $\hbos$. It can be derived from the spin-magnitude distribution at formation as follows:

\begin{equation}\label{eq:chiMdist}
    \begin{split}
	&p(\chi_{M,i} | \alpha,\beta,\mu_s,M_i,\tau_s) \\
	&=\int_0^1 p(\chi_{M,i}| \mu_s,M_i,\tau_s,\chi_{F,i}) p(\chi_{F,i}|\alpha,\beta) d\chi_{F,i},
	\end{split}
\end{equation}
in which we define the conditional probability
\begin{equation}\label{eq:chiMGivenchiF}
    \begin{split}
    &p(\chi_{M,i} |\mu_s,M_i,\tau_s, \chi_{F,i}) \\
    &=\delta\left[ \chi_{M,i}-\chi_{[nlm]}(\mu_s,M_i,\tau_s,\chi_{F,i})\right] \Theta\left[\chi_{F,i}-\chi_{[nlm]}\right] \\
    &+\delta(\chi_{M,i}-\chi_{F,i})\Theta\left[\chi_{[nlm]} - \chi_{F,i}\right],
    \end{split}
\end{equation}
where $\delta[...]$ is a Dirac delta that maps from the spin at formation to the spin at merger, and $\Theta[...]$ is a Heaviside step function that enforces the superradiance condition $\chi_{F,i}>\chi_{[nlm]}$.
Since the growth time scale $\tau^{\rm grow}_{[nlm]}$ depends mildly on the initial spin at the onset of the superradiance, the postsuperradiant spin $\chi_{[nlm]}$ depends on four parameters $(\mu_s, M, \tau_s, \chi_F)$ in principle.
One needs to calculate $\chi_{[nlm]}$ for each $\chi_{F,i}$ to precisely evaluate the integral in Eq.~\eqref{eq:chiMdist}.
To facilitate the evaluation of this integral during the sampling process, we simplify the dependence of $\chi_{F,i}$ in $\chi_{[nlm]}$ with the following approximations to fix the values of $\chi_{I[nlm]}$.
For the dominant $[011]$ mode, we set the spin at the onset to the midpoint between the minimum spin value required for superradiance and the maximal Kerr spin: $\chi_{I[011]} \approx (1+\chi_{[011]})/2$. This is justified because the growth time scale varies by only around 1 order of magnitude for spins in the range $\chi_{I[011]} \in [\chi_{[011]},1]$.
After spin-down, the BH settles on the postsuperradiant spin of the given mode, $\chi_{[nlm]}$. 
Therefore, we approximate the initial spin of each subsequent mode $\chi_{[n(l+1)(m+1)]}$ by the preceding mode postsuperradiant spin, i.e., $\chi_{I[n(l+1)(m+1)]}\approx \chi_{[nlm]}$.
With the above simplifications, $\chi_{[nlm]}$ only depends on $(\mu_s,M_i,\tau_s)$ and we can approximate Eq.~\eqref{eq:chiMdist} as
\begin{equation}
    \begin{split}
	p(\chi_{M,i} | \alpha,\beta, &\mu_s,M_i,\tau_s) \approx f_{\rm SR} \delta(\chi_{M,i}-\chi_{[nlm]}) \\
	&+ p(\chi_{M,i}|\alpha,\beta) \Theta(\chi_{[nlm]}-\chi_{M,i}),
	\end{split}
\end{equation}
with $f_{\rm SR}$ being the \textit{differential} fraction of BHs that undergo superradiance at the BH mass $M_i$:
\begin{equation}\label{eq:fsr}
\begin{split}
&f_{\rm SR}(\mu_s,M_i,\tau_s, \alpha, \beta) \\
\equiv &\int_0^1 p(\chi_{F,i} |\alpha,\beta) \Theta\left[\chi_{F,i}-\chi_{[nlm]}(\mu_s,M_i,\tau_s)\right]d\chi_{F,i}.
\end{split}
\end{equation}

In the astrophysical model, \hnull, one obtains a similar expression for $p(\Lambda_{\hnull}| \mathbf{d}, \hnull)$ by replacing $\hbos \rightarrow \hnull$ everywhere, and removing all references to $\tau_s$, which is not a relevant parameter of $\hnull$.

While calculating Eq.~\eqref{eq:hyperpos}, we use a power law prior $\pi(M_1)\propto M_1^{-2.35}$ for the primary mass~\cite{Salpeter:1955it}, and uniform prior for mass ratio $q=M_2/M_1$ with $M_2\leq M_1$.
We also assume that $\tau_s$ is known (for the \hbos model) and fixed at $10\, \rm Myr$: $\pi(\tau_s) = \delta(\tau_s -10\,{\rm Myr})$, which is toward the lower limit of the typical inspiral time scales, according to numerical simulations ($\sim 10 \, {\rm Myr}-10\, {\rm Gyr})$~\cite{2000ApJ...528L..17P,MillerLauburg:2008,2009MNRAS.395.2127O,2011MNRAS.416..133D,2012PhRvD..85l3005K,2013ApJ...777..103T,2014MNRAS.441.3703Z,2015PhRvL.115e1101R,2016PhRvL.116b9901R,2015ApJ...800....9M,dominik2013double}.
From one side, this choice is conservative as it allows for the least time for BHs to spin-down due to superradiance, making it harder to find evidence for bosons. 
From the other side, restricting the merger time prior overestimates the prior information thus overestimates the evidence for the boson hypothesis $\mathcal{H}_B$ in Eq.~\eqref{eq:hyperpos}.
Nevertheless, the additional parameter space in $\tau_s$ is expected to contribute modestly to the Bayes factor because the corrections to the available mass-spin parameter space are mostly smaller than the mass-spin measurement uncertainties.
Therefore, while more realistic models for the merger time prior $\pi(\tau_s)$ could be used, our choice is sufficient for advanced GW detectors, given their limited precision in the measurement of component masses and spins.

Since we assume the BH mass distribution is known, we do not need to consider its selection effect while calculating $\Sigma(\Lambda)$.
We can also ignore selection effects due to BH spins as the expected number of observations only varies by $\lesssim10\%$ for different spin models~\cite{farr2017distinguishing,wysocki2018hba,ng2018chieff}.
Furthermore, for $\Lambda_B$, the fraction of detectable BBHs does not depend on $\mu_s$.
Based on the above arguments, we therefore assume $\Sigma(\Lambda_{\hbos})$ and $\Sigma(\Lambda_{\hnull})$ are constants in the evaluation of the hyperposteriors.
In generating the simulations, however, we fully account for all selection effects so that the number of sources can be interpreted as the expected number of detections in future observations.

Integrating Eq.~\eqref{eq:hyperpos}, and the equivalent expression for \hnull, over the whole hyperparameters space yields evidences $Z_{\hbos}$ and $Z_{\hnull}$ that can be used to calculate the Bayes factor between the boson and astrophysical hypothesis: $\mathcal{B}_{\rm{A}}^{\rm{B}}={Z_{\hbos}}/{Z_{\hnull}}$.
We also perform a Monte Carlo simulation of 50 different sets of sources in every simulated universe.
This allows us to estimate the probability distribution of the Bayes factors due to Poisson fluctuation for each number of detections $N$.

\section{Mock data analysis}
\label{sec:result}
The method described above can be applied to both simulated and real detections. 
We first demonstrate its use on three different simulated ``universes'': (i) one with a boson scalar field with $\mu_s=10^{-13} \, \rm eV$, (ii) one with a boson with $\mu_s=10^{-12} \, \rm eV$, and (iii) one where no boson exists (``astrophysical population'').
To create the mock populations, we generate BBHs with component masses $M_{1,2}$ following the same prior in the model: $\pi(M_1)\propto M_1^{-2.35}$, uniform distribution for $q$ in $[0.1,1]$ and require both $\{M_1,M_2\} \in [5,50]\rm M_{\odot}$, consistently with Ref.~\cite{LIGOScientific:2018jsj}.
The BBHs are distributed uniformly in the source-frame comoving volume, as well as the sky positions, orbital orientations and polarization angles in the unit sphere.
The astrophysical processes that set the initial spin magnitude and orientation are still to be fully understood~\cite{Belczynski:2017gds,gerosa2018spins,postnov2019spins,Bavera:2019fkg}.
For each of the three universes, we consider two distributions of formation spin magnitudes $\chi_{F,i}$: (a) uniform in $[0,1)$ (``flat spin'') and (b) $p(\chi_{F,i})\propto (1-\chi_{F,i})$ (``low spin''), with an isotropic spin orientation in both cases.
The true shape parameters of the beta distribution are $\alpha=\beta=1$ and $\alpha=1,\beta=2$ for the ``flat spin'' and ``low spin'' populations, respectively. 

When simulating the universes where bosons exist, we need to evolve the BH spins at formation to the spins at merger using Eq.~\eqref{eq:HBSpins}. We assume all BBHs have a short merger time scale $\tau_s=10$~Myr, which minimizes the effect of superradiance and is thus a conservative choice.
To keep the computational cost of the analysis reasonable, of all the sources we generate, we only analyze those for which SNR$>30$.
These are the only sources that will contribute to the test since individual spins are hard to measure for low or medium SNR BBHs~\cite{vitale2014spinmagpe,vitale2017spinmagpe}.
The populations of synthetic BBH sources are thus added into simulated noise of the LIGO and Virgo detectors at design sensitivity~\cite{ligopsd,virgopsd}.
We use the \textsc{LALInference}~\cite{Veitch:2014wba,lalsuite} algorithm with the \textsc{IMRPhenomPv2} waveform family~\cite{smith2016fast} to obtain posterior and likelihood distributions for the compact binary parameters of the simulated sources, which can be used to infer the population hyperparameters as described in the previous section.
For all of the hyperparameters, we use uniform-in-log priors, with ranges $[0.01,10]$ for $\alpha$ and $\beta$, as well as $[10^{-13}, 3 \times 10^{-12}]$~eV for $\mu_s$, which is the range of $\mu_s$ that can be realistically probed with ground-based GW detectors~\cite{Arvanitaki:2010sy,Arvanitaki:2014wva,Arvanitaki:2016qwi,Brito:2017zvb}.

\begin{figure}[!ht]
\centering
\includegraphics[width=0.95\linewidth]{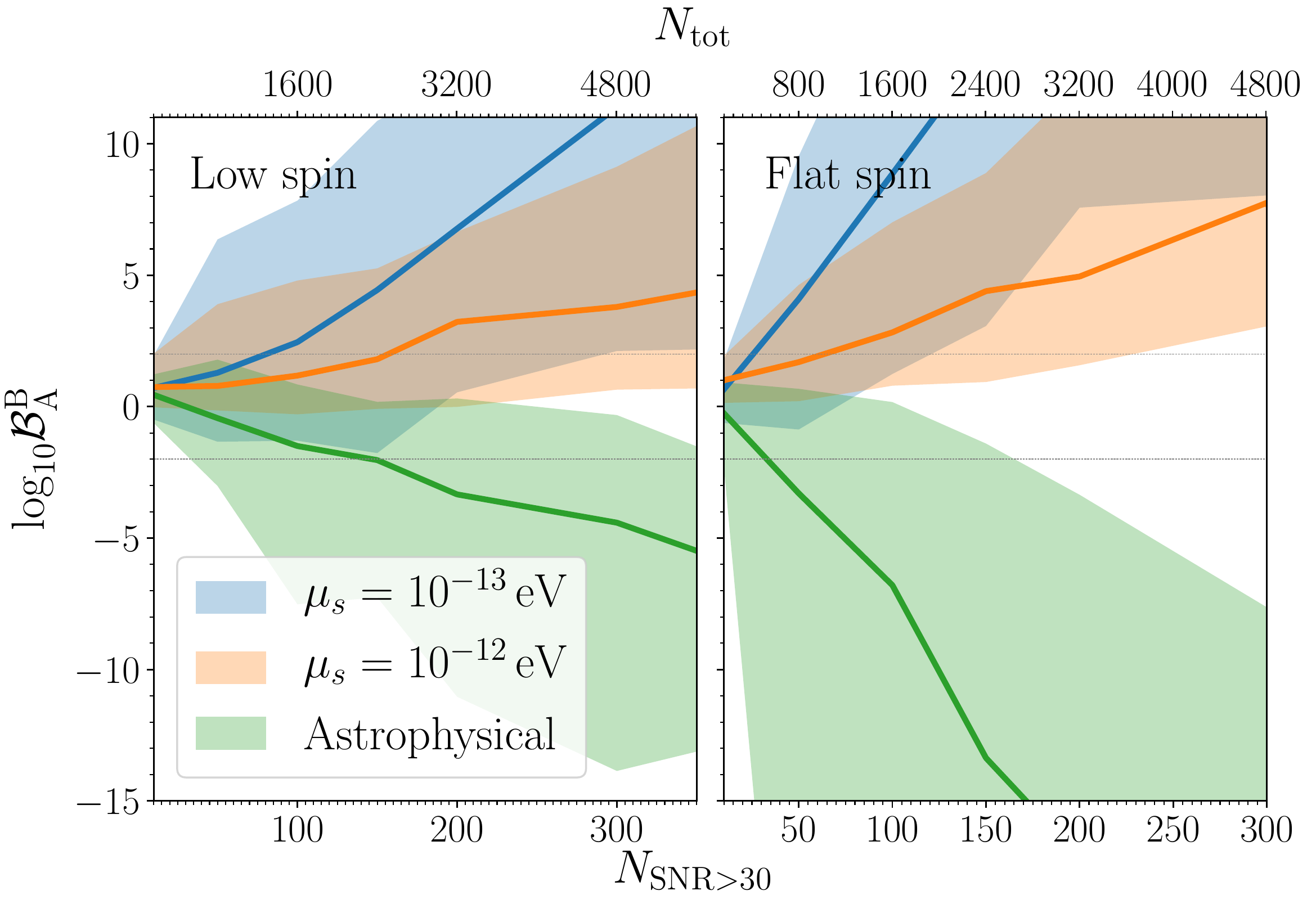}
\caption{The $\log_{10}$ Bayes factor between the boson and astrophysical hypothesis as a function of the number of sources $N_{\rm SNR>30}$ from the boson with $\mu_s=10^{-13} \, \mathrm{eV}$ (blue), boson with $\mu_s=10^{-12} \, \mathrm{eV}$ (orange), and astrophysical (green) populations.
For each population, we repeat the analysis with the low spin (left panel) and flat spin (right panel) distribution at formation.
The solid lines and colored bands are medians and 90\% credible intervals over 50 realizations of a population with $N_{\rm SNR>30}$ sources.
The two horizontal lines show $\mathcal{B}_{\rm{A}}^{\rm{B}}=0.01$ and 100.}
\label{fig:mockdataanalysis}
\end{figure}

In Fig.~\ref{fig:mockdataanalysis}, we show the evolution of the log Bayes factor boson versus astrophysical model, $\log_{10} \mathcal{B}^{\rm B}_{\rm A}$ as more events are used for the test. The bottom $x$-axes show the numbers of loud events, while the top ones show the numbers of total events.~\footnote{Since the distribution of SNRs for BBH detected by advanced detectors is known analytically and goes as $P(\rho)\propto\rho^{-4}$~\cite{Schutz:2011tw}, one can calculate that there is one event with ${\rm SNR}>30$ for each 16 events with SNR$\geq12$ on average.}

All curves show that the underlying hypothesis is correctly preferred by the method, given enough number of observations.
In Table~\ref{tab:BF}, we report the expected numbers of observations required to significantly~\footnote{We follow Ref.~\cite{ly2016harold} and strongly prefer the boson (astrophysical) hypothesis if $\mathcal{B}_{\rm{A}}^{\rm{B}}\geq 100$ ($\leq0.01$).}
prefer a hypothesis for all pairs of the spin distribution and boson mass. In general, we would expect that fewer sources are required to disprove the boson hypothesis (first row of Table~\ref{tab:BF}) than to confirm it.
This is because even one highly spinning BH measurement can contradict \hbos, whereas multiple BHs that match the predicted postsuperradiant spins are necessary to favor \hbos.
On the other hand, for some values of the boson mass, the morphology of the postsuperradiant spin distribution, including its dependence of the BH mass, may be very different from the astrophysical spin model, making it easier to prefer the boson hypothesis over the astrophysical hypothesis. This is, for example, the case for $\mu_s\sim 10^{-13}\, \rm eV$, whose exclusion region does not constrain BH masses below $\sim 15~\mathrm{M}_{\odot}$ (see Fig. 2 of ~\cite{Ng:2020ruv}). Unless the astrophysical spin distribution is significantly correlated with the BH mass (which does not seem to be the case based on the latest LVK results~\cite{GWTC2rate}), it is harder for the astrophysical model to match the expected postsuperradiant spin distribution, and hence is easier to verify the boson hypothesis for that boson mass. For heavier bosons, however, the resulting postsuperradiant spin distribution is similar to an astrophysical model (in the absence of bosons) with low BH spins at birth, which makes the two models harder to distinguish. 
This explains why smaller numbers of sources are required on average to confirm the existence of a boson with mass $\mu_s=10^{-13}\, \rm eV$ than to rule out bosons in the mass range $10^{-13} \text{ eV} \leq \mu_s\leq 3\times10^{-12} \text{ eV}$, in the flat spin and low spin scenarios, Table~\ref{tab:BF}.

We also notice that more sources are required for the test if BHs generally have low formation spins (``low spin'' population) than for the ``flat spin'' population.
This is expected since it is harder to prove the existence of a dearth of highly spinning BHs due to superradiant spin-down given a population with small formation spins.

\begin{table}
\begin{ruledtabular}
\caption{\label{tab:BF}
The estimated numbers of high SNR detections to rule out or confirm bosons for different combinations of formation spin distribution and boson.
}
\begin{tabular}{l l l}
Population models & Flat spin & Low spin \\
\hline
Astrophysical~\footnote{The statistical requirement is $\mathcal{B}^B_A=0.01$ to rule out bosons within $[10^{-13},3{\times}10^{-12}]$~eV.} & \NruleoutFlat & \NruleoutLow \\
$\mu_s=10^{-13}$~eV~\footnote{\label{confirm}The statistical requirement is $\mathcal{B}^B_A=100$ to confirm the bosons.} & \NconfirmLightFlat & \NconfirmLightLow \\
$\mu_s=10^{-12}$~eV.~\footref{confirm}& \NconfirmHeavyFlat & \NconfirmHeavyLow~\footnote{The upper bound is only an approximation since even using all the simulated signals we do not reach the desired threshold $\mathcal{B}^B_A=100$.} \\
\end{tabular}
\end{ruledtabular}
\end{table}

\begin{figure}[!ht]
\centering
\includegraphics[width=0.99\linewidth]{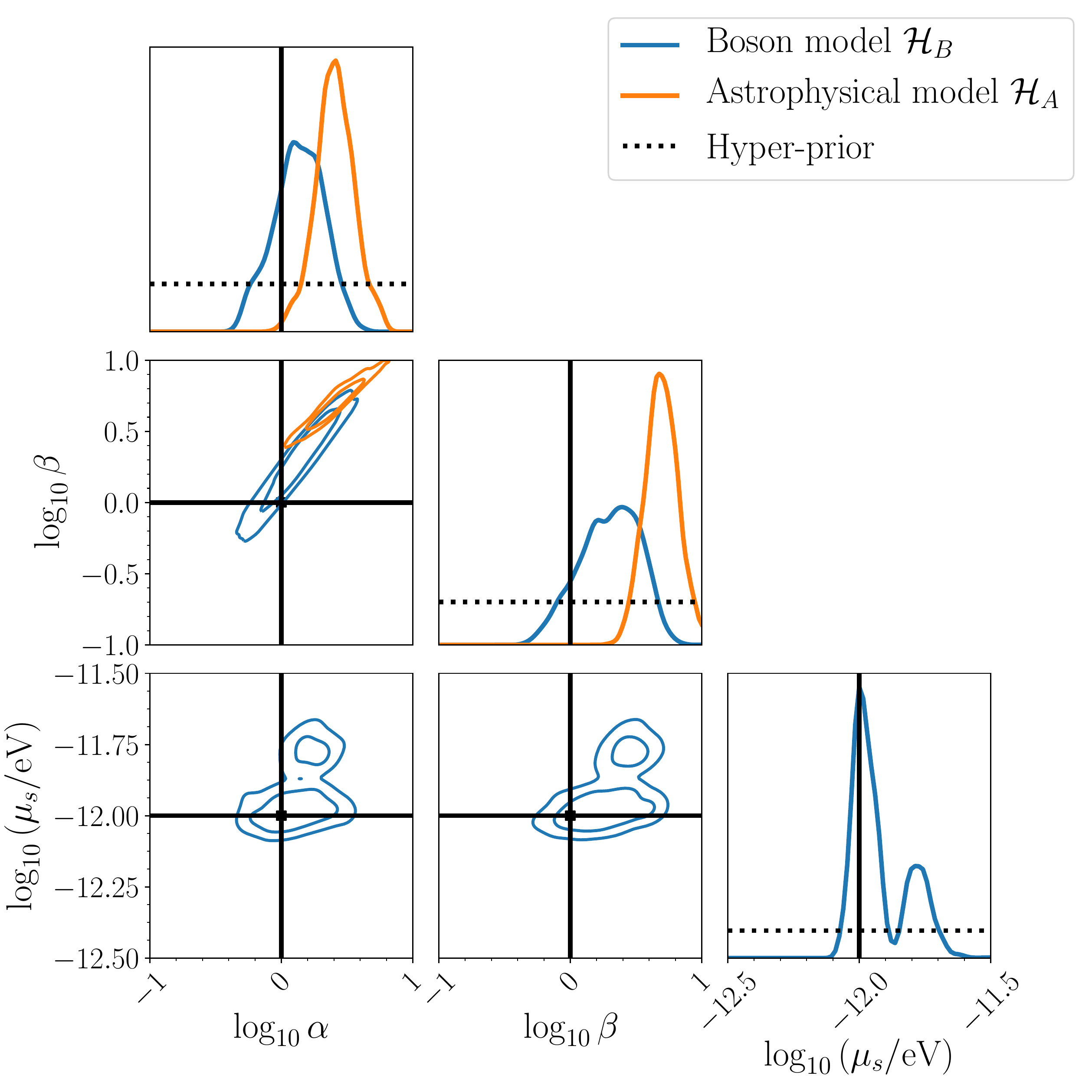}
\caption{\label{fig:hyperpos}
Corner plots of the $(\alpha,\beta,\mu_s)$ hyperposterior assuming boson model \hbos (blue) and \hnull (orange) in $\mathrm{log}_{10}$ space.
The contours are shown at 68\% and 95\% intervals.
For this example, we average 50 sets of sources, each with 300 high SNR events drawing from the boson population at $\mu_s=10^{-12}$~eV and $(\alpha,\beta)=(1,1)$ (``flat spin'').
The dashed and solid black lines are the hyperpriors and true values, respectively.
}
\end{figure}

Next, we look at the estimation of the individual hyperparameters.
As an example, we take 300 high SNR detections drawn from the simulated Universe with ``flat spin'' and $\mu_s=10^{-12}$~eV.
Figure~\ref{fig:hyperpos} shows the corner plot of $(\alpha,\beta,\mu_s)$ in $\mathrm{log}_{10}$ space, assuming \hbos (blue) or \hnull (orange).
First, we notice the model $\hbos$ results in a bimodal hyperposterior for $\mu_s$.
This is because the exclusion region generated by the \emph{first} superradiant mode of a boson with mass $\mu_s=10^{-12}$~eV is similar to the one generated by the \emph{second} mode of a boson with roughly twice the mass. In turn, this implies at least a partial degeneracy between the two configurations.
We note that the true $\mu_s$ is found at the primary peak, and the secondary peak becomes less prominent as the number of detections increases.

\begin{figure}[!ht]
\centering
\includegraphics[width=0.99\linewidth]{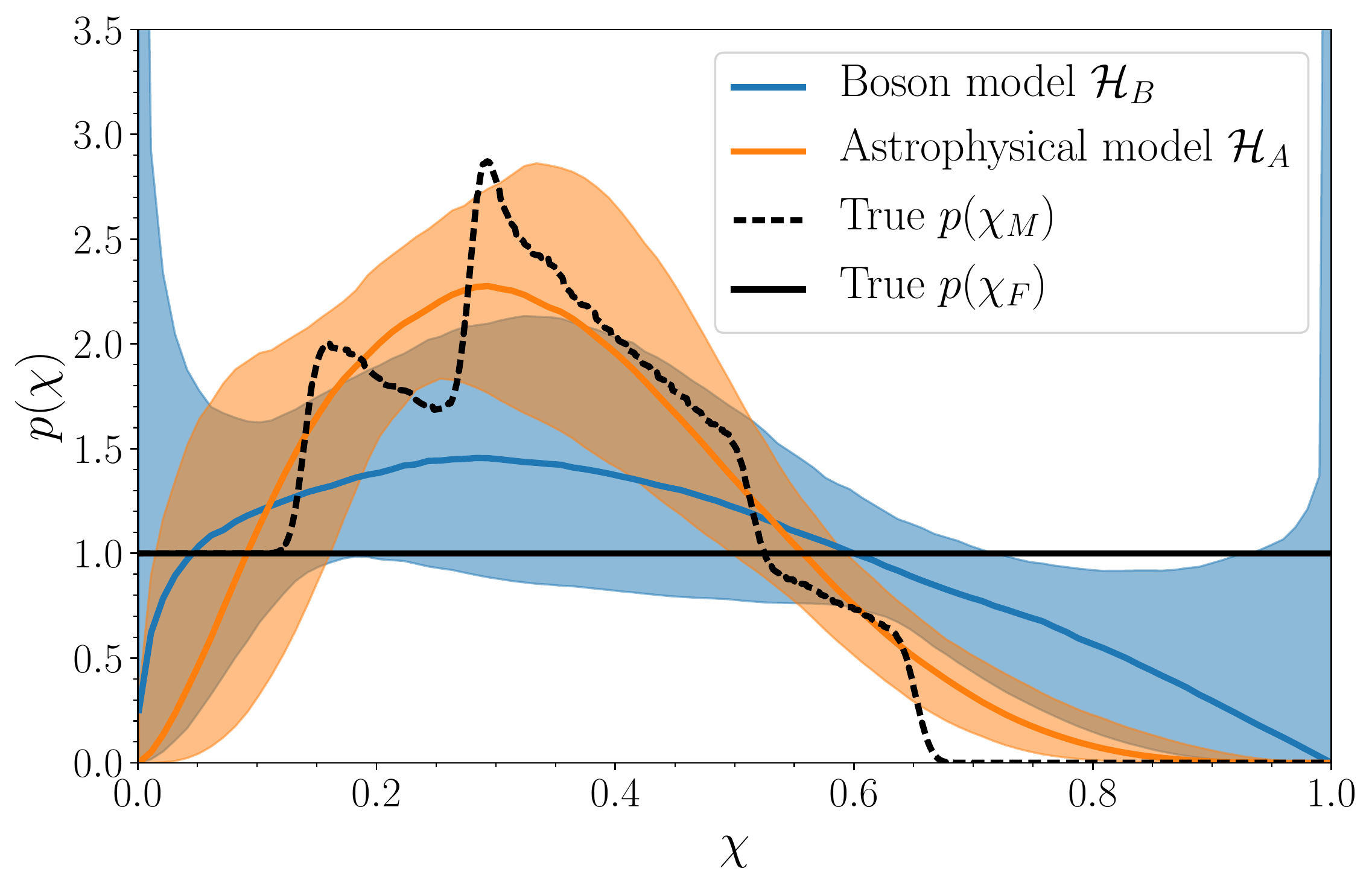}
\caption{\label{fig:poppos}
Hyperposterior of the spin distribution at formation $p(\chi_F|\alpha,\beta)$, inferred with the same set of simulations (flat spin and $\mu_s=10^{-12}$~eV) in Fig.~\ref{fig:hyperpos}.
The blue (orange) solid line shows the median of the inferred $p(\chi_F|\alpha,\beta)$ assuming \hbos (\hnull).
The bands mark the 90\% credible intervals.
The dashed and solid black lines are the true spin distribution at merger and at formation, respectively, in this simulation with boson existence.
The characteristic ``zig-zag'' structure in $p(\chi_M)$ reflects the postsuperradiant spins of different superradiant modes.
}
\end{figure}

Second, using the astrophysical model $\hnull$, we recover heavily biased values of $(\alpha,\beta)$, which control the shape of the spins at formation.
To better visualize this bias, we recast the $(\alpha,\beta)$ hyperposteriors of both models into $p(\chi_F|\alpha,\beta)$, as shown in Fig.~\ref{fig:poppos}.
The model \hnull (orange band) is indeed more consistent with the postsuperradiant spin distribution \emph{at merger} $p(\chi_M | \alpha=1, \beta=1, \mu_s=10^{-12}\,\mathrm{eV})$ (black dashed line), instead of the spin distribution at formation $p(\chi_F | \alpha=1, \beta=1)$ (black solid line).
This is not surprising, since \hnull cannot account for the superradiant spin loss and simply treats the spin at merger as if it were the spin at formation, i.e., $p(\chi_M)$ as $p(\chi_F)$.
On the other hand, \hbos (blue band) can ``undo'' the superradiance and reconstruct $p(\chi_F)$ much closer to the ``true'' distribution at formation (black solid line) in our simulation.
Hence, both the hyperposterior and the $\chi_F$ distribution, inferred by the model \hbos, are unbiased.

\section{Discussion}\label{sec:discuss}
In this paper, we have illustrated the use of hierarchical Bayesian inference to simultaneously measure the BH spin distribution at formation and to search for ultralight bosons, by combining mass and spin measurements from a population of BBHs detected by GW observatories.
Our method relies on the morphology of the expected exclusion region in the BH mass-spin plane, which depends on the properties of the boson mass, to search for and characterize the boson. 
Applying this method on a mock data set, we have shown that BBHs discovered by ground-based GW detectors can be used to rule out the existence of ultralight bosons in the mass range $[10^{-13},3\times10^{-12}]$~eV.
Our method is also capable to reveal the existence of a ultralight boson, and measure its mass, as we have explicitly shown with two example of boson masses, $\mu_s=10^{-13}$~eV and $10^{-12}$~eV.
In order to investigate the impact of spin distribution at formation on the statistical power of our method, we have generated populations of simulated BBHs with either a uniform distribution or a linearly decreasing distribution of BH spins at formation.
We found that in both cases combining $\lesssim 300$ high SNR events will be enough to rule out or confirm the existence of an ultralight boson within the mass range $[10^{-13},10^{-12}]$~eV.

While we only consider scalar bosons in this study, the method we developed is applicable to vector or tensor boson fields, which have much shorter instability and GW emission time scales~\cite{Baryakhtar:2017ngi,Cardoso:2018tly, Brito:2020lup}.
Our analysis of simulated BBHs has made a few simplifying assumptions which make it conservative. First, we have assumed that all BBHs merge in 10~Myr, which is toward the lower limit of what is usually obtained in numerical simulations~\cite{2000ApJ...528L..17P,MillerLauburg:2008,2009MNRAS.395.2127O,2011MNRAS.416..133D,2012PhRvD..85l3005K,2013ApJ...777..103T,2014MNRAS.441.3703Z,2015PhRvL.115e1101R,2016PhRvL.116b9901R,2015ApJ...800....9M,dominik2013double}.
Since most of the BBHs in our parameters space of interest would have undergone superradiance within 10~Myr except for the very low mass systems $\sim 5M_{\odot}$, assuming longer merger times does not significantly improve the searching efficiency.
Second, we have assumed that only sources with SNR$>30$ will contribute to this test, as their spins are easier to measure.
In reality, while the component spins of weaker events are harder to measure, they will still contribute to the test.

In the analysis, we ignored the $\lesssim10\%$ BH mass loss during superradiance.
In order to assess the impact of this choice, we repeated the analysis with all BH mass posteriors shifted by $5\%$ toward the light side, hence mimicking the effect of mass loss.
This translates to a systematic overestimation of $\sim5\%$ boson mass, which is still within the statistical uncertainty, in our inference with $\mathcal{O}(100)$ high SNR sources.
However, we expect this systematic error will dominate when the number of events grow to $\mathcal{O}(10^4)$ in the era of next-generation detectors~\cite{Punturo:2010zz,InstrumentSciencePaper,Vitale:2018yhm,LIGOScientific:2019vkc,Maggiore:2019uih,Reitze:2019iox,Adhikari:2019zpy,Hall:2019xmm}.

The true distribution of spins at formation plays the most important role: the number of events needed to perform this test will be larger if the astrophysical distribution of spins at formation is such that small spins are preferred. Conversely, if many highly spinning BHs are formed, potentially with significant misalignments between spin and angular momenta (both of which make spins easier to measure), then fewer sources will be necessary.
Given the measured BBH merger rate, ground-based interferometers will detect hundreds of BBHs per year at design sensitivity~\cite{Dominik:2014yma,Ng:2017yiu,Oguri:2018muv,Baibhav:2019gxm,Aasi:2013wya}.
In this large-number observations regime, one will want to use more sophisticated models which also capture eventual correlations between the masses and spins of astrophysical BHs~\cite{Belczynski:2017gds,gerosa2018spins,Bavera:2019fkg}.
This may boost or suppress the statistical power of testing boson hypothesis in a different boson mass range, depending on the actual joint distribution of BH mass and spin at formation.
We will leave investigating these systematics to future work.

Within the assumptions made in this study, it seems feasible to rule out the existence of ultralight bosons \emph{everywhere} in the mass range of $[10^{-13},3\times10^{-12}]$~eV with a few years of advanced detectors data.
However, we note that fewer GW events would be required to rule out the bosons in a \emph{narrower} boson mass range. 
Statistically proving the existence of these bosons will take longer, as more sources are required: the planned upgrades of LIGO and Virgo to their ``plus'' configurations might yield thousands of BBH events per year, which will make it more plausible to gather the evidence for the existence of ultralight bosons in the mass range of $[10^{-13},3\times10^{-12}]$~eV~\cite{InstrumentSciencePaper,LIGOScientific:2019vkc}.

\section{Acknowledgements}
We thank the anonymous referees for their suggestions which significantly improved this paper.
We also thank Emanuele Berti, Richard Brito, Will Farr, Carl Haster, Max Isi, and Kaze Wong for the valuable discussions and suggestions.
K. K. Y. N. and S. V. acknowledge the support of the National Science Foundation (NSF) through the NSF Grant No. PHY-1836814. LIGO was constructed by the California Institute of Technology and Massachusetts Institute of Technology with funding from the National Science Foundation and operates under cooperative agreement PHY-1764464.
The work of O. A. H. was supported by the Hong Kong Ph.D. Fellowship Scheme issued by the Research Grants Council of Hong Kong before resubmission and was supported by the research program of the Netherlands Organization for Scientific Research.
The work of T. G. F. L. was partially supported by grants from the Research Grants Council of Hong Kong (Projects No. CUHK14306218, No. CUHK14310816, and No. CUHK24304317), Research Committee of the Chinese University of Hong Kong, and the Croucher Foundation in Hong Kong.
This research made use of data, software, and/or web tools obtained from the Gravitational Wave Open Science Center~\cite{Abbott:2019ebz}, a service of LIGO Laboratory, the LIGO Scientific Collaboration, and the Virgo Collaboration. The authors are grateful for computational resources provided by the LIGO Lab and supported by the NSF Grants No. PHY-0757058 and No. PHY-0823459.

\bibliography{paper}
\end{document}